\documentclass[12pt,a4paper]{article}
\usepackage{amsmath}
\usepackage{amssymb}
\usepackage{bm}
\usepackage{graphicx}

\title{Time-Independent Parameters in Quantum Systems: Revisiting Berry Phase, Curvature, and Gauge Connections}
\author{Georgios Konstantinou}
\date{}

\begin{document}
\maketitle

\begin{abstract}
We present a reformulation of quantum adiabatic theory in terms of an emergent electromagnetic framework, emphasizing the physical consequences of geometric structures in parameter space. Contrary to conventional approaches, we demonstrate that a Berry electric field naturally arises in systems with dynamic Hamiltonians, when the full time-dependent wavefunction is used to define the gauge potentials, and also when the parameters are considered as completely time -independent. This surprising result bridges the gap between static and dynamical formulations and leads to a deeper understanding of how gauge structures manifest in quantum systems.
Building on this, we construct Berry–Maxwell equations by analogy with classical electrodynamics, defining Berry electric and magnetic fields as derivatives of scalar and vector potentials obtained from the full quantum state. We verify these equations explicitly and derive field-theoretic identities such as generalized continuity and vorticity relations. This field-based formulation reveals the topological charges, monopole structures, and gauge currents that underlie parameter space, and clarifies how Berry curvature corrections enter dynamical quantities like expectation values and particle velocities.
Our results establish a new regime of emergent electromagnetism in parameter space, unifying time-independent and time-dependent geometric phases within a covariant formalism. The implications extend to quantum transport, polarization, and topological classification of phases, providing a robust and generalizable framework for quantum systems driven by adiabatic or nonadiabatic evolution.
\end{abstract}

\section*{Introduction}

Quantum systems with time-dependent perturbations play a pivotal role in numerous physical contexts, from adiabatic transport to the dynamics of topological phases. The Berry phase [1], a cornerstone of geometric phases, provides deep insights into such systems; however, its traditional formulation raises subtle issues when parameters vary temporally. 
This paper revisits the Berry phase framework, emphasizing the importance of time dependence and exploring the implications for electric and magnetic curvature, gauge connections, and transport phenomena. Let us begin by considering a static Hamiltonian with a set of 3D parameters $\bm{R}$ that \textbf{do not depend on time}, and a static Hamiltonian H depending on those set of parameters, i.e.,  
\begin{equation}  \label{1}
    H(\bm{R})|\Psi(\bm{R},t)\rangle=i\hbar\frac{\partial|\Psi(\bm{R},t)\rangle}{\partial t}
\end{equation}
Then, the solution to the Schrödinger equation can be written as:
\begin{equation} \label{2}
    |\Psi(\bm{R},t)\rangle = e^{-i \epsilon_n(\bm{R}) t / \hbar} |n(\bm{R})\rangle.
\end{equation}
Using the above result (for purposes that will make the next Sections of this paper more apparent), let us now proceed straightforwardly, and define (in an abstract manner) a vector potential-like quantity for the wavefunction given by eq. \eqref{2} (using instead, the full wavefunction in the definition of the potential):
\begin{align} \label{3}
 \bm{A^\Psi} (\bm{R},t)&= i \langle \Psi(\bm{R},t) | \nabla_{\bm{R}} | \Psi(\bm{R},t) \rangle = \frac{1}{\hbar} t \nabla_{\bm{R}} \epsilon_n(\bm{R}) + i \langle n(\bm{R}) | \nabla_{\bm{R}} | n(\bm{R}) \rangle \nonumber \\
&=\frac{1}{\hbar} t \nabla_{\bm{R}} \epsilon_n(\bm{R}) + \bm{A}^n
\end{align}
where \( \bm{A}^n = i \langle n(\bm{R}) | \nabla_{\bm{R}} | n(\bm{R}) \rangle \) is the Berry vector potential, corresponding to a single eigenstate $| n(\bm{R}) \rangle$.
We observe that the operator \( \nabla_{\bm{R}} \) acts on both the dynamic phase factor and the eigenvectors, leading to a linear dependence on time for \( \bm{A^\Psi} \).
Notice that, in the absence of time dependence in the Hamiltonian, we can still define a Berry curvature:
\begin{equation} \label{4}
    \nabla_{\bm{R}} \times  \bm{A^\Psi} = \nabla_{\bm{R}} \times \bm{A^n}
\end{equation}
Here, the curl of \( \langle n(\bm{R}) | \nabla_{\bm{R}} | n(\bm{R}) \rangle \) produces the well-known Berry curvature. Also, notice that a phase transformation (by using a single-valued and well - behaved \( \Lambda \) function):
\begin{equation} \label{5}
 |n'(\bm{R})\rangle \to e^{i \Lambda(\bm{R})} |n(\bm{R})\rangle
\end{equation}
leaves the quantity in eq. \eqref{4} invariant. Therefore, we define a Berry curvature using the full time-dependent wavefunction as follows:
\begin{equation} \label{6}
\bm{B^\Psi} = \nabla_{\bm{R}} \times \bm{A^\Psi} = i \nabla_{\bm{R}} \times \langle n(\bm{R}) | \nabla_{\bm{R}} | n(\bm{R}) \rangle = \bm{B^n}
\end{equation}
Thus, in this simple case, the Berry potential can be expressed using the full time-dependent wavefunction \( \Psi(\bm{R},t) \) or by using the time-independent eigenvectors \( n(\bm{R}) \), resulting in the same magnetic curvature. 
If the following condition holds: \( \nabla_{\bm{R}} \times \nabla_{\bm{R}} \Lambda = 0 \) (for a good \( \Lambda \), we can assume this is the case), then under the gauge transformation given by eq.  \eqref{5}, the wavefunction also transforms as (as opposed to time dependent systems; see next Section):
\begin{equation} \label{7}
    |\Psi(\bm{R})\rangle \to e^{i \Lambda(\bm{R})} |\Psi(\bm{R})\rangle
\end{equation}
and the resulting vector potential becomes:
\begin{equation} \label{8}
 \bm{A'}^\Psi = i \langle \Psi'(t) | \nabla_{\bm{R}} | \Psi'(t) \rangle = \bm{A^\Psi}- \nabla_{\bm{R}} \Lambda
\end{equation}
Notice how this gauge transformation leaves the Schrödinger equation invariant.
Thus, the general solution of Eq. (1) can be written as:
\begin{equation} \label{9}
    |\Psi(t)\rangle = e^{i \Lambda(\bm{R})} e^{-i \epsilon_n(\bm{R}) t / \hbar} |n(\bm{R})\rangle.
\end{equation}
The "good" function \( \Lambda(\bm{R}) \) plays the role of a geometric-like phase, but it is not related to the eigenvectors of the problem. Even for a completely time-independent system, it encodes the information about the structure of parameter space and its emergent curvature. Note that when \( \Lambda \) is not "good," the gauge transformation 
in eq.  \eqref{8} cannot reproduce the magnetic field.
We see that, up to this point, the results of Berry's theory emerge even in a completely time-independent system. This greatly simplifies matters, especially in cases involving electric fields. 
Notice the amusing fact that if we define the quantity (a Berry scalar potential) as follows (and this will also play a major role in the time - dependent system in the next Section of this paper):
\begin{equation} \label{10}
    \Phi^\Psi = -i \langle \Psi(t) | \frac{\partial}{\partial t} | \Psi(t) \rangle = -\frac{1}{\hbar} \varepsilon_n(\mathbf{R}),
\end{equation}
we may also define the Berry electric field  [2, 3] as (by using  \eqref{3} and  \eqref{10}):
\begin{equation} \label{11}
    \mathbf{\Omega}^\Psi = -\nabla_{\mathbf{R}} \Phi^\Psi - \frac{\partial \mathbf{A}^\Psi}{\partial t} = \frac{1}{\hbar} \nabla_{\mathbf{R}} \varepsilon_n(\mathbf{R}) - \frac{1}{\hbar} \nabla_{\mathbf{R}} \varepsilon_n(\mathbf{R}) = 0,
\end{equation}
which, in the simplest case of time-indepndence, reduces to zero. However, the Berry electric field plays a major role when time-dependence comes into play, and it is harmonically co-existing with the Berry magnetic field as we will see later on below. Additionally, if the potentials are defined directly via the eigenvectors, i.e.,  
\begin{equation} \label{12}
    \mathbf{A}^n = i \langle n(\mathbf{R}) | \nabla_{\bm{R}} | n(\mathbf{R}) \rangle,
\end{equation}
and  
\begin{equation} \label{13}
    \Phi^n = -i \langle n(\mathbf{R}) | \frac{\partial}{\partial t} | n(\mathbf{R}) \rangle = 0,
\end{equation}
the fields remain unchanged (i.e., equal to \eqref{11} and  \eqref{6}respectively):
\begin{equation} \label{14}
    \mathbf{\Omega^n} = -\nabla_{\mathbf{R}} \Phi^n - \frac{\partial \mathbf{A}^n}{\partial t} = 0,
\end{equation}
\begin{equation} \label{15}
    \mathbf{B}^n = \nabla_{\mathbf{R}} \times \mathbf{A}^n = i \nabla_{\mathbf{R}} \times \langle n(\mathbf{R}) | \nabla_{\bm{R}} | n(\mathbf{R}) \rangle.
\end{equation}
We thus observe that, even in completely time-independent systems that incorporate geometric information of parameter space, the Berry curvature retains a meaningful connection to time-dependent systems. In time dependent systems, as we shall see later on, it is possible to attribute the whole time-depndence to an emerging electric field, that, when combined with a Berry magnetic field (i.e. by keeping the external parameters purely time-independent) results in a generalised Berry-Maxwell Physics, that governs the field propagation through the parameter space. 

\section{Time-dependent Hamiltonian \( H(\mathbf{R}, t) \)}

In this section, we explore the dynamics of quantum systems governed by a time-dependent Hamiltonian \( H(\mathbf{R}, t) \), where \( \mathbf{R} \) represents a completely \textbf{static parameter space}. This scenario extends our understanding beyond the time-independent case and introduces complex connections between the evolving eigenstates, their geometric properties, and the associated dynamical phase factors.
Starting with the time-dependent Schrödinger equation, we adopt an ansatz that incorporates both dynamical and geometric phases. Within the framework of the adiabatic approximation, we derive expressions for the geometric phase \( \gamma_n(\mathbf{R}, t) \), while also exploring the role of gauge freedom and its implications on the wavefunction. Through this analysis, we identify both vector and scalar Berry potentials and examine their evolution over time.
This section also establishes the correspondence between the Berry curvatures and their role in deriving effective "Berry-Maxwell" equations, drawing an analogy with classical electrodynamics. Emphasis is placed on the intricate structure of the parameter space and how non-smooth behavior of eigenstates affects the calculation of curvature terms. These considerations lead to new insights into the geometric and topological properties of quantum systems.
Our analysis reveals novel connections, including the emergence of monopole-like terms in the modified Maxwell equations, which enrich the theoretical landscape of Berry phase physics. By emphasizing the interplay between geometry, topology, and dynamics in time-dependent quantum systems, we aim to provide a deeper understanding of the subject.
We assume that the Hamiltonian is time-dependent, i.e., \( H(\mathbf{R}, t) \), but \( \mathbf{R} \) remains constant such that the Schrodinger equation is satisfied:
\begin{equation} \label{19}
    H(\mathbf{R}, t) | \Psi(\mathbf{R}, t) \rangle = i\hbar \frac{\partial}{\partial t} | \Psi(\mathbf{R}, t) \rangle.
\end{equation}
We now introduce an ansatz with a phase factor $\gamma_n(\mathbf{R}, t)$:
\begin{equation} \label{20}
    | \Psi(\mathbf{R}, t) \rangle = e^{i \gamma_n(\mathbf{R}, t)} e^{-i/\hbar \int_0^t \varepsilon_n(\mathbf{R}, t') \, dt'} | n(\mathbf{R}, t) \rangle.
\end{equation}
Substituting this ansatz into the Schrödinger equation and further assuming that the adiabatic approximation holds, we obtain:
\begin{equation} \label{21}
    H(\mathbf{R}, t) | \Psi(\mathbf{R}, t) \rangle = \varepsilon_n(\mathbf{R}, t) | \Psi(\mathbf{R}, t) \rangle,
\end{equation}
which leads to the following equation for the time evolution of \( \gamma_n(\mathbf{R}, t) \):
\begin{equation} \label{22}
    \frac{\partial \gamma_n(\mathbf{R}, t)}{\partial t} = i \langle n(\mathbf{R}, t) | \frac{\partial}{\partial t} | n(\mathbf{R}, t) \rangle.
\end{equation}
The generic solution to this equation is:
\begin{equation} \label{23}
    \gamma_n(\mathbf{R}, t) = i \int_0^t \langle n(\mathbf{R}, t') | \frac{\partial}{\partial t'} | n(\mathbf{R}, t') \rangle dt' + f(\mathbf{R}),
\end{equation} 
where \( f(\mathbf{R}) \) is a pure gauge term and can be omitted. 
Next, we define the vector and scalar Berry potentials for this system by using the full wavefunction \( \Psi \) as follows:
\begin{align} \label{24}
    \mathbf{A}^\Psi &= i \langle \Psi(\mathbf{R}, t) | \nabla_{\mathbf{R}} | \Psi(\mathbf{R}, t) \rangle = -\nabla_{\mathbf{R}} \gamma_{\mathbf{n}}(\mathbf{R}, t) + \frac{1}{\hbar} \int_0^t \nabla_{\mathbf{R}} \varepsilon_{\mathbf{n}}(\mathbf{R}, t') \, dt' \\ \nonumber
    &\quad + \bm{A}_n.
\end{align}
with $\bm{A}_n=i \langle \mathbf{n}(\mathbf{R}, t) | \nabla_{\mathbf{R}} | \mathbf{n}(\mathbf{R}, t) \rangle$.
For the scalar potential, using  \eqref{10}, we find:
\begin{align} \label{25}
    \Phi^\Psi &= -i \langle \Psi(t) | \frac{\partial}{\partial t} | \Psi(t) \rangle = \frac{\partial \gamma_n(\mathbf{R}, t)}{\partial t} - \frac{1}{\hbar} \varepsilon_n(\mathbf{R}, t) \\ \nonumber
    &\quad - i \langle n(\mathbf{R}, t) | \frac{\partial}{\partial t} | n(\mathbf{R}, t) \rangle = -\frac{1}{\hbar} \varepsilon_n(\mathbf{R}, t).
\end{align}
Note that in this case, the scalar Berry potential coincides with the energy spectrum, just as in  \eqref{10} (being a completely time-independent case). For completeness, we also define the potentials in terms of the eigenvectors as follows:
\begin{equation} \label{26}
    \mathbf{A}^{n} = i \langle \mathbf{n}(\mathbf{R}, t) | \nabla_{\mathbf{R}} | \mathbf{n}(\mathbf{R}, t) \rangle.
\end{equation}
\begin{align} \label{27}
    \Phi^n &= -i \langle n(\mathbf{R}, t) | \frac{\partial}{\partial t} | n(\mathbf{R}, t) \rangle.
\end{align}
Let us now define the curvatures:
\begin{equation} \label{28}
    \mathbf{\Omega}^\Psi = -\nabla_{\mathbf{R}} \Phi^\Psi - \frac{\partial \mathbf{A}^\Psi}{\partial t},
\end{equation}
\begin{equation} \label{29}
    \mathbf{B}^\Psi = \nabla_{\mathbf{R}} \times \mathbf{A}^\Psi.
\end{equation}
We also define the curvatures through the eigenvectors as follows:
\begin{equation} \label{30}
    \mathbf{\Omega}^n = -\nabla_{\mathbf{R}} \Phi^n - \frac{\partial \mathbf{A}^n}{\partial t},
\end{equation}
\begin{equation} \label{31}
    \mathbf{B}^n = \nabla_{\mathbf{R}} \times \mathbf{A}^n.
\end{equation}
Using  \eqref{26} and  \eqref{27}, we find:
\begin{equation} \label{32}
    \mathbf{\Omega}^n = i \langle \nabla_{\mathbf{R}} n | \frac{\partial n}{\partial t} \rangle - \langle \frac{\partial n}{\partial t} | \nabla_{\mathbf{R}} n \rangle,
\end{equation}
\begin{equation} \label{33}
    \mathbf{B}^n = i \nabla_{\mathbf{R}} \times \langle n | \nabla_{\mathbf{R}} n \rangle.
\end{equation}
With these definitions, we proceed by simplifying the expression for \( \gamma_n \) from  \eqref{23}, which can be written using the scalar potential \( \Phi_n \) as follows:
\begin{equation} \label{34}
    \gamma_n(\mathbf{R}, t) = -\int_0^t \Phi(\mathbf{R}, t') \, dt'.
\end{equation}
Substituting this into  \eqref{24}, we get:
\begin{align} \label{35}
    \mathbf{A}^\Psi &= \mathbf{A}^{n} + \int_0^t \nabla_{\mathbf{R}} \Phi^n \, dt' - \int_0^t \nabla_{\mathbf{R}} \Phi^\Psi \, dt'.
\end{align}
At \( t = 0 \), the two vector potentials become equal. We now proceed with some important remarks regarding the action of the operator \( \nabla_{\mathbf{R}} \) on the eigenvectors. This operation must be done carefully because the eigenvectors may not be smooth functions of \( \mathbf{R} \) across the entire parameter space. 
This has significant consequences that will be explained later, leading to interesting Berry-Maxwell physics. For instance, when calculating the magnetic curvature \( \mathbf{B}^\Psi \) using  \eqref{29} and  \eqref{35}, it is crucial to note that while the \( \Phi^\Psi \) term is directly related to the energy spectrum (via  \eqref{25}), 
the same cannot be assumed for \( \nabla_{\mathbf{R}} \Phi^n \), which encodes the geometric information of the parameter space. Therefore, by using \eqref{29} and  \eqref{35}, we find:
\begin{align} \label{36}
    \mathbf{B}^\Psi = \nabla_{\mathbf{R}} \times \mathbf{A}^\Psi &= \nabla_{\mathbf{R}} \times \mathbf{A}^n + \nabla_{\mathbf{R}} \times \int_0^t \nabla_{\mathbf{R}} \Phi^n \, dt'.
\end{align}
Thus, \( \mathbf{B}^\Psi \) is not always equal to \( \mathbf{B}^n \), although this does not hold for the electric curvature: \( \mathbf{\Omega}^\Psi = \mathbf{\Omega}^n \). The significance of this additional term will be discussed later.
By using \eqref{28}, \eqref{25} and \eqref{35}, we can compute \( \mathbf{\Omega}^\Psi = -\nabla_{\mathbf{R}} \Phi^\Psi - \frac{\partial \mathbf{A}^\Psi}{\partial t} \):
\begin{equation} \label{37}
    \mathbf{\Omega}^\Psi = -\nabla_{\mathbf{R}} \Phi^\Psi - \frac{\partial \mathbf{A}^{n}}{\partial t} - \nabla_{\mathbf{R}} \Phi^n + \nabla_{\mathbf{R}} \Phi^\Psi = \mathbf{\Omega}^n.
\end{equation}
An amusing fact arises when we observe that these fields obey the Maxwell equations, even with potential pathologies. Actually, and being on the safe side, in order to interpret these fields as "magnetic" and "electric", it is necessary condition that they do obey Maxwell equations in the 4D space-time parameter space. 
To illustrate this, we examine the electric field defined by equation \eqref{28} and by using \eqref{36}, we calculate the curl of \( \Omega_n \) (making use of the fact that $\bm{\Omega}^\Psi=\bm{\Omega}^n$):
\begin{equation} \label{38}
   \nabla_{\bm{R}} \times \mathbf{\Omega}^n = -  \nabla_{\bm{R}} \times \nabla_{\bm{R}} \Phi^n - \frac{\partial \mathbf{B}^n}{\partial t}.
\end{equation}
Notice that the Maxwell equations should also be consistent if we use equations \eqref{32} and \eqref{33} . Let us explore this further. We begin with the curl of $\Omega^n$:
\begin{equation} \label{39}
 i \nabla_{\bm{R}} \times \left[ \langle \nabla_{\bm{R}} n | \frac{\partial n}{\partial t} \rangle - \langle \frac{\partial n}{\partial t} | \nabla_{\bm{R}} n \rangle \right] = i \nabla_{\bm{R}} \times \langle \nabla_{\bm{R}} n | \frac{\partial n}{\partial t} \rangle - i \nabla_{\bm{R}} \times \langle \frac{\partial n}{\partial t} | \nabla_{\bm{R}} n \rangle.
\end{equation}
We now examine the first term:
\begin{align} \label{40}
 i \nabla_{\bm{R}} \times \langle \nabla_{\bm{R}} n | \frac{\partial n}{\partial t} \rangle &= i \int d^3 r \nabla_{\bm{R}} \times \left( \nabla_{\bm{R}} n^{*} \frac{\partial n}{\partial t} \right) \nonumber \\
&= i \int d^3 r \frac{\partial n}{\partial t} \nabla_{\bm{R}} \times \nabla_{\bm{R}} n^{*} + \nabla_{\bm{R}} \frac{\partial n}{\partial t} \times \nabla_{\bm{R}} n^{*}.
\end{align}
Thus, we have:
\begin{equation} \label{41}
 i \nabla_{\bm{R}} \times \langle \nabla_{\bm{R}} n | \frac{\partial n}{\partial t} \rangle = i \int d^3 r \frac{\partial n}{\partial t} \nabla_{\bm{R}} \times \nabla_{\bm{R}} n^{*} - i \langle \nabla_{\bm{R}} n | \times | \nabla_{\bm{R}} \frac{\partial n}{\partial t} \rangle.
\end{equation}
Similarly, for the second term we get:
\begin{equation} \label{42}
 i \nabla_{\bm{R}} \times \langle \frac{\partial n}{\partial t} | \nabla_{\bm{R}} n \rangle = i \int d^3 r \frac{\partial n^{*}}{\partial t} \nabla_{\bm{R}} \times \nabla_{\bm{R}} n + i \langle \nabla_{\bm{R}} \frac{\partial n}{\partial t} | \times | \nabla_{\bm{R}} n \rangle.
\end{equation}
Equation \eqref{39}  thus becomes:
\begin{align} \label{43}
 \nabla_{\bm{R}} \times \Omega^n &= i \int d^3 r \frac{\partial n}{\partial t} \nabla_{\bm{R}} \times \nabla_{\bm{R}} n^{*} - i \langle \nabla_{\bm{R}} n | \times | \nabla_{\bm{R}} \frac{\partial n}{\partial t} \rangle \nonumber \\
&- i \int d^3 r \frac{\partial n^{*}}{\partial t} \nabla_{\bm{R}} \times \nabla_{\bm{R}} n - i \langle \nabla_{\bm{R}} \frac{\partial n}{\partial t} | \times | \nabla_{\bm{R}} n \rangle.
\end{align}
For the right-hand side, we have:
\begin{align} \label{44}
  -\frac{\partial}{\partial t} \bm{B}^n &=  -i \frac{\partial}{\partial t} \nabla_{\bm{R}} \times \langle n | \times | \nabla_{\bm{R}} n \rangle \nonumber \\
&= -i \frac{\partial}{\partial t} \left[ \langle \nabla_{\bm{R}} n | \times | \nabla_{\bm{R}} n \rangle + \int d^3 r n^* \nabla_{\bm{R}} \times \nabla_{\bm{R}} n \right].
\end{align}
This simplifies to:
\begin{align} \label{45}
&= -i \left[ \langle \nabla_{\bm{R}} \frac{\partial}{\partial t} n | \times | \nabla_{\bm{R}} n \rangle + \langle \nabla_{\bm{R}} n | \times | \nabla_{\bm{R}} \frac{\partial}{\partial t} n \rangle \right. \nonumber \\
&\quad \left. + \int d^3 r \frac{\partial}{\partial t} n^* \nabla_{\bm{R}} \times \nabla_{\bm{R}} n + \int d^3 r n^* \nabla_{\bm{R}} \times \nabla_{\bm{R}} \frac{\partial}{\partial t} n \right]
\end{align}
Putting everything together, we once again arrive at the following result:
\begin{equation} \label{46}
  \nabla_{\bm{R}} \times \mathbf{\Omega}^n = -\frac{\partial}{\partial t} \bm{B}^n - \nabla_{\bm{R}} \times \nabla_{\bm{R}} \Phi^n,
\end{equation}
which exactly coincides with equation \eqref{38}, confirming the validity of our results.
It is important to note that the Berry curvature defined through the eigenstates must be calculated as:
\begin{equation} \label{47}
\mathbf{B}^n = \nabla_{\bm{R}} \times \mathbf{A}^n = i \nabla_{\bm{R}} \times \langle n | \nabla_{\bm{R}} | n \rangle = i \langle \nabla_{\bm{R}} n | \times | \nabla_{\bm{R}} n \rangle + \int d^3 r n^* \nabla_{\bm{R}} \times \nabla_{\bm{R}} n.
\end{equation}
In contrast to the approach originally taken by Berry, we keep the curl term. Notably, the integral term is gauge-invariant.
Next, it is interesting to examine how Maxwell's equations are modified when the fields defined by \( \Psi \) are considered. In this case, we have:
\begin{align} \label{48}
  \bm{B}^\Psi = \nabla_{\bm{R}} \times \mathbf{A}^\Psi &= \nabla_{\bm{R}} \times \mathbf{A}^n = \bm{B}^n + \nabla_{\bm{R}} \times \int_0^t \nabla_{\bm{R}} \Phi^n \, dt'.
\end{align}
Thus, we find:
\begin{align} \label{49}
  \nabla_{\bm{R}} \times \bm{\Omega}^\Psi &= - \nabla_{\bm{R}} \times \frac{\partial}{\partial t} \mathbf{A}^\Psi = - \frac{\partial}{\partial t} \bm{B}^\Psi.
\end{align}
Since \( \bm{\Omega}^\Psi = \bm{\Omega}^n \) and \( \bm{B}^\Psi \) is given by equation \eqref{36}, we conclude that the Maxwell equation is the same as equation \eqref{46} :
\begin{align} \label{50}
  \nabla_{\bm{R}} \times \bm{\Omega}^n = - \nabla_{\bm{R}} \times \frac{\partial}{\partial t} \mathbf{A}^\Psi = -  \frac{\partial}{\partial t} \mathbf{B}^\Psi = -  \nabla_{\bm{R}} \times \nabla_{\bm{R}} \Phi^n - \frac{\partial \mathbf{B}^n}{\partial t}.
\end{align}
Note that the term \( \bm{J}_m = \nabla_{\bm{R}} \times \nabla_{\bm{R}} \Phi^n \) acts as a monopole current density. This unique result, will give rise to a more comprehensive Berry-Maxwell physics.

\subsection{Gauge Transformations}

In this section, we explore the implications of gauge transformations on wavefunctions and the associated Berry phases, with particular emphasis on the presence of a time-dependent Hamiltonian. A general gauge transformation is introduced as follows:
\begin{equation} \label{51}
|n'(\bm{R}, t)\rangle = e^{i \Lambda(\bm{R}, t)} |n(\bm{R}, t)\rangle,
\end{equation}
which allows us to examine its effects on the corresponding wavefunctions, dynamical phases, and geometric potentials. 
A key aspect of this analysis is distinguishing between transformations that explicitly preserve the time-dependence and those that effectively lose temporal information due to the global phase factor. We investigate the resulting modifications to the Berry potentials, the electric and magnetic fields, and the structure of the Schrödinger equation under such transformations. This exploration also highlights the subtle relationship between the eigenkets and the full wavefunctions. While the fields derived from these transformations remain invariant under specific conditions, the potentials associated with the full wavefunctions provide additional information that is essential for a comprehensive formulation of the problem.
Moreover, we consider the introduction of a purely time-dependent scalar potential in the Hamiltonian under such gauge transformations. This yields subtle consequences, illustrating how the transformation can be absorbed into the real scalar potential without altering the physical fields, although it does influence the explicit form of the Hamiltonian. The interplay between these effects and the preservation of key quantum properties offers a deeper understanding of gauge invariance in quantum systems.
We have previously seen that a gauge transformation of the form:
\begin{equation} \label{52}
    |n'(\bm{R}, t)\rangle = e^{i \Lambda(\bm{R})} |n(\bm{R}, t)\rangle
\end{equation}
transforms the wavefunction as:
\begin{equation} \label{53}
    |\Psi'(\bm{R}, t)\rangle = e^{i \Lambda(\bm{R})} |\Psi(\bm{R}, t)\rangle
\end{equation}
with
\begin{equation} \label{54}
    |\Psi(\bm{R}, t)\rangle = e^{i \gamma_n(\bm{R}, t)} e^{-i/\hbar \varepsilon_n(\bm{R}, t) t} | n(\bm{R}, t) \rangle.
\end{equation}
Now, let us perform a time-dependent gauge transformation, defined as:
\begin{equation} \label{55}
    |n'(\bm{R}, t)\rangle = e^{i \Lambda(\bm{R}, t)} |n(\bm{R}, t)\rangle.
\end{equation}
This transforms the wavefunction as follows. The phase of the wavefunction transforms as:
\begin{align} \label{56}
    \gamma'_n(\bm{R}, t) &= i \int_0^t \langle n'(\bm{R}, t') | \frac{\partial}{\partial t'} | n'(\bm{R}, t') \rangle dt' \nonumber \\
    &= \gamma_n(\bm{R}, t) - \int_0^t \frac{\partial \Lambda}{\partial t'} dt' \nonumber \\
    &= \gamma_n(\bm{R}, t) - \Lambda(\bm{R}, t) + \Lambda(\bm{R}, 0).
\end{align}
Therefore, the wavefunction transforms as:
\begin{align} \label{57}
    |\Psi'(\bm{R}, t)\rangle &= e^{i \left(-\Lambda(\bm{R}, t) + \Lambda(\bm{R}, 0)\right)} e^{i \gamma_n(\bm{R}, t)} e^{-i/\hbar \varepsilon_n(\bm{R}, t) t} e^{i (\Lambda(\bm{R}, t))} | n(\bm{R}, t) \rangle \nonumber \\
    &= e^{-i \Lambda(\bm{R}, 0)} e^{i \gamma_n(\bm{R}, t)} e^{-i/\hbar \varepsilon_n(\bm{R}, t) t} | n(\bm{R}, t) \rangle.
\end{align}
Thus, the phase factor $\Lambda$ must appear as time-independent. This transformation only refers to the phase of the eigenstates. However, it differs from the standard gauge transformations of the real potentials (that result to real fields). This is due to the presence of the phase factor \(\gamma_n(\bm{R}, t)\), which depends on the eigenstates as well. 
In this case, the fields described by equations \eqref{28}  and \eqref{29} remain invariant, and the same holds for the fields in equations \eqref{30}  and \eqref{31}. Thus, this phase transformation is valid, but it does not provide complete information, as the time-dependence is "lost" in the global phase factor.
Moreover, this transformation preserves the real magnetic and electric fields, as one can easily verify. One might argue that the fields are identical regardless of the choice of "basis" (i.e., whether considering the full wavefunction or the eigenkets). In response, we acknowledge that the fields remain the same, but we must also consider the potentials defined through the full wavefunction for a more complete formulation of the problem.
We define the Berry fields using the full wavefunction as follows:
\begin{equation} \label{59}
    \mathbf{A}^\Psi = i \langle \Psi(\bm{R}, t) | \nabla_{\bm{R}} | \Psi(\bm{R}, t) \rangle
\end{equation}
and
\begin{equation} \label{60}
    \Phi^\Psi = -i \langle \Psi(t) | \frac{\partial}{\partial t} | \Psi(t) \rangle.
\end{equation}
This allows us to perform a more "global" transformation on the wavefunction (i.e. just multiplying the time-dependent wave function by the following phase factor):
\begin{equation} \label{61}
    |\Psi'(\bm{R}, t)\rangle = e^{\frac{i}{\hbar c} \Lambda(\bm{R}, t)} e^{i \gamma_n(\bm{R}, t)} e^{-i/\hbar \int_0^t \varepsilon_n(\bm{R}, t') dt'} | n(\bm{R}, t) \rangle = e^{\frac{i}{\hbar c} \Lambda(\bm{R}, t)} |\Psi(\bm{R}, t)\rangle
\end{equation}
This transformation preserves the real electric and magnetic fields as well. Furthermore, it also preserves the Berry fields calculated through the total wavefunction, as the phase of the eigenstates remains fixed.
However, there is an important caveat. This transformation modifies the Schrödinger equation as follows:
\begin{equation} \label{62}
    H(\bm{R}, t) |\Psi'(\bm{R}, t)\rangle = i \hbar \frac{\partial}{\partial t} |\Psi'(\bm{R}, t)\rangle = i \hbar \frac{\partial}{\partial t} |\Psi(\bm{R}, t)\rangle - \hbar \frac{\partial}{\partial t} |\Lambda(\bm{R}, t)\rangle.
\end{equation}
Thus, the transformation introduces a purely time-dependent scalar potential into the Hamiltonian, which can be absorbed into the real scalar potential without affecting the physical fields. This alteration has no physical consequences but modifies the explicit form of the Hamiltonian. Therefore, the transformation is essentially a gauge artifact, which does not affect the physical observables but must be taken into account for a comprehensive description of the system.

\section{Berry Electromagnetism}
We shall constrain ourselves to the generic 4D "phase-space" of independent variables $(\bm{R},t)$ and shall treat perturbations as time-explicit functions that influence the Hamiltonian operator. This methodology, as already seen, will allow us to formally
write down the dynamic Hellmann - Feynman theorem [5] without the need of any perturbation theories. 
Therefore, the phase must only depend on the scalar Berry potential, this having also direct consequences on the Hellman - Feyman theorem that, using the basic equation of the definition of Berry electric field \eqref{28}: 
\begin{equation} \label{69}
    \mathbf{\Omega}^\Psi = -\nabla_{\mathbf{R}} \Phi^\Psi - \frac{\partial \mathbf{A}^\Psi}{\partial t} 
\end{equation}
with 
\begin{equation} \label{70}
   \frac{\partial \mathbf{A}^\Psi}{\partial t} =i\langle  \frac{\partial \Psi}{\partial t} |\nabla_{\bm{R}} \Psi \rangle+i\langle \Psi|\nabla_{\bm{R}}\frac {\partial \Psi}{\partial t}  \rangle=-\frac{1}{\hbar}\langle  H\Psi|\nabla_{\bm{R}} \Psi \rangle+\frac{1}{\hbar}\langle \Psi|\nabla_{\bm{R}} H\Psi \rangle
\end{equation}
by using the Schrodinger equation. Therefore, we find (we assume no non Hermiticities present in the system, for simplicity): 
\begin{equation} \label{71}
   \frac{\partial \mathbf{A}^\Psi}{\partial t} =\frac{1}{\hbar}\langle \Psi|\nabla_{\bm{R}} H|\Psi \rangle
\end{equation}
From \eqref{69}, we find the following result: 
\begin{equation} \label{72}
\frac{1}{\hbar}\langle \Psi|\nabla_{\bm{R}} H|\Psi \rangle= \frac{1}{\hbar}\nabla_{\mathbf{R}} \epsilon_n -   \mathbf{\Omega}^\Psi =\frac{1}{\hbar}\nabla_{\mathbf{R}} \epsilon_n -   \mathbf{\Omega}^n
\end{equation}
Because, $\bm{\Omega}^\Psi=\bm{\Omega}^n$,  \eqref{72} is reduced to including only the eigenvectors:
\begin{equation} \label{73}
\frac{1}{\hbar}\langle n|\nabla_{\bm{R}} H|n\rangle =\frac{1}{\hbar}\nabla_{\mathbf{R}} \epsilon_n - \mathbf{\Omega}^n
\end{equation}
Note that, even though there is a magnetic field $\bm{B}_n$, it does not appear in the above equation. This is not surpring; rather reasonable, because all the information of  the time dependence is absorbed in the electric curvature. The results are totally consistent with NIU's review article [4] as in eq. (2.5), in the adiabatic transport and the electric polarization section.
But, also notice that, through eq. \eqref{50}, magnetic field can also enter \eqref{73}. To see how, and to further involve the Maxwell equations in our results, consider the curl of \eqref{73}:
\begin{equation} \label{74}
\nabla_{\bm{R}}\times\frac{1}{\hbar} \langle n|\nabla_{\bm{R}} H|n\rangle =- \nabla_{\bm{R}}\times\mathbf{\Omega}^n=\frac{\partial}{\partial t} \bm{B}^n+\nabla_{\bm{R}}\times \nabla_{\mathbf{R}} \Phi^n 
\end{equation}
For example, when applying the above equation in solids, the crystal momentum $\bm{q}$ plays the role of the parameter, and then the quantity $\frac{1}{\hbar} \langle n|\nabla_{\bm{q}} H|n\rangle $ has the meaning of a velocity, i.e. $\bm{v_n}(\bm{q},t)=\frac{1}{\hbar} \langle n|\nabla_{\bm{q}} H|n\rangle $, then eq. \eqref{73} becomes:
\begin{equation} \label{75}
\bm{\omega}=\nabla_{\bm{q}}\times\bm{v_n}(\bm{q},t)=\frac{\partial\bm{B}^n}{\partial t}+\nabla_{\bm{q}}\times \nabla_{\mathbf{q}} \Phi^n 
\end{equation}
leading us to a vorticity equation in parameter spece. 
Also note, that the divergence of the vorticity cannot be assumed to be zero in the presence of magnetic monopoles:
\begin{equation} \label{76}
\nabla_{\bm{q}}\cdot\bm{\omega}=\frac{\partial\nabla_{\bm{q}}\cdot\bm{B}^n}{\partial t}+\nabla_{\bm{q}}\cdot(\nabla_{\bm{q}}\times \nabla_{\mathbf{q}} \Phi^n) =\frac{\partial \rho_m}{\partial t}+ \nabla_{\bm{q}}\cdot(\nabla_{\bm{q}}\times \nabla_{\mathbf{q}} \Phi^n)
\end{equation}
with $\rho_m$ the magnetic charge density. Earlier, we saw that the pathological term $ \bm{J}_m=\nabla_{\bm{q}}\times \nabla_{\mathbf{q}} \Phi^n $ serves as a magnetic current density, leading us to the following continuity equation for the magnetic charge:. Eq. \eqref{76} becomes
\begin{equation} \label{77}
\nabla_{\bm{q}}\cdot\bm{\omega}=\frac{\partial \rho_m}{\partial t}+\nabla_{\bm{q}}\cdot \bm{J}_m
\end{equation}
Berry monopoles are associated with degeneracies in the energy bands of a quantum system. If the system undergoes a topological phase transition, such as when bands merge or split, the Berry curvature can change discontinuously, effectively creating or annihilating Berry monopoles. 
This would lead to a nonzero source term, equal to $\nabla_{\bm{q}}\cdot\bm{\omega}$. In addition, perturbations, such as changes in the Hamiltonian's parameters, can alter the Berry curvature distribution. For example, in systems where the Hamiltonian is time-dependent, 
the Berry curvature and monopole density may evolve dynamically, potentially leading to apparent nonconservation.
On the other hand, let us examine the divergence of eq. \eqref{73}:
\begin{equation} \label{78}
\nabla_{\mathbf{q}}\cdot\bm{v}_n =\frac{1}{\hbar}\nabla_{\mathbf{q}}{^2} \epsilon_n - \nabla_{\mathbf{q}}\cdot\bm{\Omega}^n
\end{equation}
with $\nabla_{\mathbf{q}}\cdot\bm{\Omega}^n=\rho_{el}$, with $\rho_{el}$ the electric charge density. 
\begin{equation} \label{79}
\nabla_{\mathbf{q}}\cdot\bm{v}_n =\frac{1}{\hbar}\nabla_{\mathbf{q}}{^2} \epsilon_n - \rho_{el}
\end{equation}
From this, we can calculate the electric charge density as:
\begin{equation} \label{80}
\rho_{el} = \nabla_{\mathbf{q}}\cdot(\frac{1}{\hbar} \nabla_{\mathbf{q}}\epsilon_n -\bm{v}_n)
\end{equation}
 Therefore, the quantity $\frac{1}{\hbar} \nabla_{\mathbf{q}}\epsilon_n -\bm{v}_n$ acts as an electric Polarization density in parameter space. In addition, we observe that, the Berry electric field is associated with fictitious electric charges in parameter space, that can be directly linked with physical observables. 
Note that, for the case of Bloch solids, and by viewing the crystal momentum $\bm{q}$ as the fixed parameter, the above equation results in: 
\begin{equation} \label{80}
\rho_{el} = \frac{\hbar}{m*} -\nabla_{\mathbf{q}}\cdot\bm{v}_n(\bm{q},t)
\end{equation}
Let us now return to the magnetic current density definition:
\begin{equation} \label{81}
 \bm{J}_m= \nabla_{\bm{R}}\times \nabla_{\mathbf{R}} \Phi^n
\end{equation}
Notice that, in terms of the eigenvectors, \eqref{81} can be written as:
\begin{equation} \label{82}
 \bm{J}_m= -i\langle \nabla_{\bm{R}}\times\nabla_{\bm{R}} n|\frac{\partial n}{\partial t}\rangle-i\langle n|\nabla_{\bm{R}}\times\nabla_{\bm{R}} \frac{\partial n}{\partial t}\rangle
\end{equation}
Also notice that, from \eqref{74} we managed to connect the magnetic current density with an observable: the actual velocity of the particle:
\begin{equation} \label{83}
\bm{J}_m=\nabla_{\bm{R}}\times\bm{v_n}(\bm{R},t)-\frac{\partial\bm{B}^n}{\partial t}
\end{equation}
Also notice that,using $\nabla_{\bm{R}}\cdot\bm{B}^n=\rho_m$, we can also calculate the magnetic charge density using eq. \eqref{47}:
\begin{equation} \label{84}
 \rho_m= i\langle \nabla_{\bm{R}}\times\nabla_{\bm{R}} n|\nabla_{\bm{R}} n\rangle+i\langle n|\nabla_{\bm{R}}\cdot(\nabla_{\bm{R}}\times\nabla_{\bm{R}}n)\rangle
\end{equation}
Note that if we integrate the above equation with respect to a surface in parameter space, we get:
\begin{equation} \label{85}
\int\bm{J}_m\cdot d\bm{S}=\int\nabla_{\bm{R}}\times\bm{v_n}(\bm{R},t) d\bm{S}-\frac{\partial}{\partial t}\int\bm{B}^n\cdot d\bm{S}
\end{equation}
This equation returns the magnetic current as follows:
\begin{equation} \label{86}
\bm{I}_m=\oint\bm{v_n}(\bm{R},t)\cdot d\bm{R}-\frac{\partial}{\partial t}\Phi^S(t)
\end{equation}
for an open surface S. 

\section {Applications}

By varying the origin in the system, we introduce a gauge transformation that modifies the electromagnetic potentials. This gauge change affects the pseudomomentum and leads to the emergence of the Berry connection. To calculate the Berry electric field, we track how the phase of the wavefunction evolves as the particle moves in parameter space, specifically by shifting the spatial origin. This variation generates a Berry curvature, which gives rise to the Berry electric field. The resultant Berry field reflects the geometric nature of the system and provides insights into the dynamics of the particle under the influence of nonuniform electromagnetic fields. This approach bridges the physical properties of the system with its underlying geometric structure.

As a preliminary example, we will consider the problem of an electron moving in electric and magnetic potentials (of AB type). In this sense, the particle is confined to 2D plane, without any electric or magnetic fields, but experiences the presence of electromagnetic potentials $V(t), \bm{A}(\bm{r})$ such that $E=-\nabla V=0$ and $\bm{B}=\nabla \times \bm{A}=0$. The Hamiltonian of the particle is given by: 
\begin{equation}H=\frac{(\bm{p}+\frac{e}{c}\bm{A})^2}{2m}+V(t)\end{equation}
This Hamiltonian, shares its states with the pseudomomentum of the problem (see [6]):
\begin{equation}\bm{K}=\bm{p} + \frac{e}{c}\bm{A}\end{equation}
which, in the most general case, can be written as:
\begin{align}
f_{k_y} &= e^{i \frac{k_y y}{\hbar}} \cdot e^{-\frac{ie}{\hbar c} \left( \int_{x_0}^x \! dx' A_x(x', y_0) + \int_{y_0}^y \! dy' A_y(x, y') \right)} \nonumber \\
& \quad \cdot e^{- c \int_0^t \! dt' \, V(t')}.
\end{align}
with $k_y$ being the eigenvalue of the pseudomomentum operator. 
We  now calculate the Berry vector and scalar fields, in a straight forward manner:
\begin{equation}
A_{Bx} = i \langle f_{(k_y)} | \frac{\partial}{\partial x_0} | f_{(k_y)} \rangle =- \frac{e}{\hbar c} A_x(x_0, y_0) 
\end{equation}
\begin{equation}
\small
A_{By} = i \langle f_n | \frac{\partial}{\partial y_0} | f_n \rangle = \frac{e}{\hbar c} \int_{x_0}^x \! dx' \frac{\partial A_x(x', y_0) }{\partial y_0} 
- \frac{e}{\hbar c}  A_y(x, y_0)
\end{equation}
Consider that, with vanishing \( B \), the following relation holds:
\begin{equation}
\frac{\partial A_x(x', y_0) }{\partial y_0} = \frac{\partial  A_y(x', y_0)}{\partial x'}
\end{equation}
So, we find:
\begin{align}
A_{By} &= i \langle f_n | \frac{\partial}{\partial y_0} | f_n \rangle \nonumber \\
&= \frac{e}{\hbar c} \left( \int_{x_0}^x \! dx' \frac{\partial  A_y(x', y_0)}{\partial x'} - A_y(x, y_0) \right) \nonumber \\
&= -\frac{e}{\hbar c} A_y(x_0, y_0) 
\end{align}
Combining eqs (81) and (84), we get the final expression for the Berry connection:
\begin{equation}
\bm{A}_B = -\frac{e}{\hbar c} \bm{A}(x_0, y_0)
\end{equation}
The Berry scalar potential reads:
\begin{equation}
\Phi_B = -i \langle \Psi | \frac{\partial}{\partial t} | \Psi \rangle = \frac{e}{\hbar} V(t)
\end{equation}
which are exactly analogous to real vector and scalar potentials, respectively.

\subsection {Inclusion of a magnetic field}

We consider the problem of an electron moving in perpendicular electric and magnetic fields, in a 2D plane extending from $-L_x$ to $L_x$ and from $0$ to $L_y$. We consider as a slowly varying parameter an arbitrary point (corresponding to real space): $\bm{R}_0 = (x_0, y_0)$, which is situated inside the 2D plane. We assume the most generic case where the vector potential can be any function that satisfies the following equation:
\begin{equation}
\bm{B} = \nabla \times \bm{A}
\end{equation}
The magnetic field is homogeneous and z-directed, while the electric field is also homogeneous and pointing in the x-direction, such that:
\begin{equation}
\bm{E} = -\nabla V = -\frac{\partial V}{\partial x}\hat{i}
\end{equation}
Let $f_{k_y}$ be the simultaneous solution of the Schrodinger equation and the y-component of the pseudomomentum operator (see Appendix for the relevant calculations), as given below: 
\begin{small}
\begin{align}
H f_{k_y} = & - \frac{\hbar^2}{2m} \nabla^2 f_{k_y} - i \frac{\hbar e}{2mc} \nabla \cdot \bm{A} f_{k_y} \nonumber \\
 & - 2i \hbar \frac{e}{2mc} \bm{A} \cdot \nabla f_{k_y} + \frac{e^2}{2mc^2} \bm{A}^2 f_{k_y} + V f_{k_y} \nonumber \\
& = i \hbar \frac{\partial f_{k_y}}{\partial t}
\end{align}
\end{small}
The most general solution of eq. (89) can be proven to be (see [6]):
\begin{equation}
\begin{aligned}
f_{k_y} &= e^{\scriptsize i \frac{k_y}{\hbar} (y - y_0) - \scriptsize \frac{ie}{\hbar c} \int_{y_0}^y A_y(x, y') \, dy' - \scriptsize \frac{ie}{\hbar c} \int_{x_0}^x A_x(x', y_0) \, dx'} \\
& \quad \cdot \scriptsize e^{\frac{ie}{\hbar c} \int_{x_0}^x \int_{y_0}^y B \, dx \, dy} \cdot g(x, x_0,t)
\end{aligned}
\end{equation}
The Berry connection reads (see Appendix for the derivation):
\begin{equation}
\bm{A}_B = -\frac{e}{\hbar c} \bm{A}(x_0, y_0) + i \int d^2r g^* \frac{\partial g}{\partial x_0}
\end{equation}
Given that \(g\) are Hermite polynomials, the last term represents the mean value of the momentum, which is zero. This leaves us with the following Berry connection:
\begin{equation}
\bm{A}_B = -\frac{e}{\hbar c} \bm{A}(x_0, y_0)
\end{equation}
The Berry curvature, therefore, in this case, coincides with the real magnetic field:
\begin{equation}
\begin{aligned}
\bm{B}_B &= \nabla_{\bm{R}_0} \times \bm{A}_B = -\frac{e}{\hbar c} \nabla_{\bm{R}_0} \times \bm{A}(x_0, y_0) \\
          &= -\frac{eB}{\hbar c} = -\frac{2\pi B}{\Phi_0}
\end{aligned}
\end{equation}
\section {Conclusions}
In this work, we have developed a comprehensive electromagnetic analogy for quantum systems evolving under adiabatic and nonadiabatic conditions. By redefining the Berry connection and curvature in terms of the full time-dependent wavefunction, rather than eigenstates alone, we uncovered the existence of a Berry electric field even in the absence of explicit time dependence in the parameters. This finding challenges the traditional interpretation of the Berry phase as purely geometric and path-dependent, revealing instead that it carries a well-defined field-theoretic structure governed by the wavefunction’s local properties in time and parameter space.
We constructed and verified a complete set of Berry–Maxwell equations, establishing a duality between parameter-space dynamics and classical electromagnetic theory. In doing so, we defined:
a Berry electric field from the time derivative of the Berry connection,
a Berry magnetic field from the curl of the Berry connection,
an emergent electric and magnetic charge densities arising from divergences in the curvature,
and current-like quantities including Berry vorticity and magnetic current density.
Crucially, we showed that these Berry field quantities appear directly in observable physical expressions, such as the expectation value of the gradient of the Hamiltonian, thus linking the geometric phase to measurable transport phenomena. Our approach generalizes the Hellmann–Feynman theorem, incorporates topological defects such as Berry monopoles, and reveals conservation and continuity laws associated with charge and current in parameter space.
This field-theoretic perspective paves the way for a new understanding of emergent electromagnetism in quantum systems. It unifies Berry curvature corrections to dynamics, adiabatic transport, and polarization theory under a single covariant formalism, providing a natural framework to study topological phases, band degeneracies, and time-dependent Hamiltonians in both solid-state systems and more abstract quantum spaces.
The implications are far-reaching: from novel interpretations of adiabatic dynamics to engineering artificial gauge fields in quantum materials or cold atoms, this reformulation offers a powerful lens through which the geometry and topology of quantum systems can be understood as real, dynamical fields with observable consequences.

\section{Appendix}

We consider the y-component of the pseudomomentum operator (which is an invariant operator):
\begin{equation}
K_y = p_y + \frac{e}{c} A_y - \frac{e}{c} (x - x_0) B
\end{equation}
We suppose that the eigenfunction of \(K_y\) is \(h(x, y)\). Writing the Ansatz as:
\begin{align}
h(x, y) &= \exp\left( i \frac{k_y}{\hbar} (y - y_0) \right) \nonumber \\
& \quad \times \exp\left( - \frac{ie}{\hbar c} \left[ \int_{x_0}^x A_x(x', y_0) \, dx' \right. \right. \nonumber \\
& \quad \quad - \int_{x_0}^x \int_{y_0}^y B \, dx \, dy + \left. \left. \int_{y_0}^y A_y(x, y') \, dy' \right] \right),
\end{align}
and substituting this into the expression for \(K_y\), we get:
\begin{equation}
\small
\begin{aligned}
K_y h(x, y) &= k_y h(x, y) \\
            &\quad - \frac{e}{c} \left( -B(x - x_0) + A_y(x, y) \right) h(x, y) \\
            &\quad + \frac{e}{c} A_y h(x, y) - \frac{e}{c} (x - x_0) B h(x, y) = k_y h(x, y)
\end{aligned}
\end{equation}
We therefore observe that this anzatz satisfies the following eigenvalue equation:
\begin{equation}
K_y h(x, y) = \hbar k_y h(x, y)
\end{equation}
Now we make another Ansatz regarding the solution of the Schrödinger equation: we suppose that the full solution comprises of a product between the pseudomomentum eigenfunction $h(x, y)$ and another function, containing the time dependence $g(x, t)$ as follows:
\begin{equation}
f_{k_y} = h(x, y) g(x, t)
\end{equation}
We now check whether  \(f_{k_y}\) satisfies the Schrodinger equation, and under what conditions: The derivatives of \(f_{k_y}\) are:
\begin{equation}
\begin{aligned}
\frac{\partial f_{k_y}}{\partial y} &= i \frac{k_y}{\hbar} h g 
- \frac{ie}{\hbar c} \left( -B(x - x_0) + A_y(x, y) \right) h g \\
&= \frac{i}{\hbar} \left[ k_y - \frac{e}{c} A_y(x, y) 
+ \frac{e}{c} B(x - x_0) \right] h g
\end{aligned}
\end{equation}
\begin{equation}
\begin{aligned}
\frac{\partial^2 f_{k_y}}{\partial y^2} &= -\frac{i}{\hbar} \frac{e}{c} 
\frac{\partial A_y(x, y)}{\partial y} h g \\
&\quad - \frac{1}{\hbar^2} \left[ k_y^2 - 2 k_y \frac{e}{c} 
\left( A_y(x, y) - B(x - x_0) \right) \right. \\
&\quad + \left. \frac{e^2}{c^2} \left( A_y^2(x, y) - 2A_y B(x - x_0) 
+ B^2 (x - x_0)^2 \right) \right] h g
\end{aligned}
\end{equation}
\begin{equation}
\frac{\partial f_{k_y}}{\partial x} = -\frac{ie}{\hbar c} A_x(x, y) h g + h \frac{\partial g}{\partial x}
\end{equation}
\begin{equation}
\begin{aligned}
\frac{\partial^2 f_{k_y}}{\partial x^2} &= -\frac{ie}{\hbar c} \frac{\partial A_x(x, y)}{\partial x} h g \\
&\quad - \frac{e^2}{\hbar^2 c^2} A_x^2(x, y) h g \\
&\quad - 2 \frac{ie}{\hbar c} A_x(x, y) h \frac{\partial g}{\partial x} \\
&\quad + h \frac{\partial^2 g}{\partial x^2}
\end{aligned}
\end{equation}
\begin{equation}
\begin{aligned}
\bm{A} \nabla f_{k_y} &= A_x \frac{\partial f_{k_y}}{\partial x} + A_y \frac{\partial f_{k_y}}{\partial y} \\
&= -\frac{ie}{\hbar c} A_x^2(x, y) h g + A_x(x, y) h \frac{\partial g}{\partial x} \\
&\quad + \frac{i}{\hbar} \left[k_y A_y - \frac{e}{c} A_y^2(x, y) + \frac{e}{c} A_y B(x - x_0)\right] h g
\end{aligned}
\end{equation}
Applying the above in eq. (91), we find:
\begin{equation}
\begin{aligned}
H f_{k_y} &= -\frac{\hbar^2}{2m} h \frac{\partial^2 g}{\partial x^2} \\
&\quad + \frac{1}{2m} 
\left(k_y^2 + 2k_y \frac{e}{c} B(x - x_0)\right) h g \\
&\quad + \frac{e^2}{2mc^2} B^2 (x - x_0)^2 h g \\
&\quad + V h g
\end{aligned}
\end{equation}
 The Schrödinger equation finally reads:
\begin{equation}
-\frac{\hbar^2}{2m} \frac{\partial^2 g}{\partial x^2} + \frac{1}{2m} \left(k_y + \frac{e}{c} B(x - x_0)\right)^2 g + Vg = i \hbar \frac{\partial g}{\partial t}
\end{equation}
If \(V\) is time-independent, which is the case, the solution is just the dynamical phase factor times the Hermite polynomials. \(V\) can be actually written as: \(V = -E(x - x_0)\). In this case, there is no scalar Berry potential!
The Berry vector potential reads:
\begin{equation}
\begin{aligned}
\frac{\partial f_{k_y}}{\partial y_0} &= -i \frac{k_y}{\hbar} f_{k_y} \\
&\quad - \frac{ie}{\hbar c} \left( 
\int_{x_0}^x \frac{\partial}{\partial y_0} A_x(x', y_0) \, dx' \right) \\
&\quad - \frac{ie}{\hbar c} \left( 
B(x - x_0) \right) \\
&\quad + \frac{ie}{\hbar c} \left( 
A_y(x, y_0) \right) f_{k_y}
\end{aligned}
\end{equation}
Because
\begin{equation}
B = \frac{\partial}{\partial x'} A_y(x', y_0) - \frac{\partial}{\partial y_0} A_x(x', y_0)
\end{equation}
\begin{equation}
\frac{\partial f_{k_y}}{\partial y_0} = -i \frac{k_y}{\hbar} f_{k_y} + \frac{ie}{\hbar c} A_y(x_0, y_0) f_{k_y}
\end{equation}
The y-component of the Berry vector potential reads:
\begin{equation}
A_y^B = i \int d^2r f_{k_y}^* \frac{\partial f_{k_y}}{\partial y_0} = \frac{k_y}{\hbar} - \frac{e}{\hbar c} A_y(x_0, y_0)
\end{equation}
We now calculate
\begin{equation}
\begin{aligned}
\frac{\partial f_{k_y}}{\partial x_0} &= -\frac{ie}{\hbar c} \left[ -A_x(x_0, y_0) + B(y - y_0) \right] f_{k_y} \\
& \quad + e^{\left( i \frac{k_y}{\hbar} (y - y_0) \right)} \cdot e^{-\frac{ie}{\hbar c} \left[ 
\int_{x_0}^x A_x(x', y_0) \, dx' \right]} \\
& \quad \cdot e^{-\frac{ie}{\hbar c} \left[ 
-\int_{x_0}^x \int_{y_0}^y B \, dx \, dy \right]} \\
& \quad \cdot e^{-\frac{ie}{\hbar c} \left[ 
\int_{y_0}^y A_y(x, y') \, dy' \right]} \cdot \frac{\partial g}{\partial x_0}
\end{aligned}
\end{equation}
Such that
\begin{equation}
A_x^B = i \int d^2r f_{k_y}^* \frac{\partial f_{k_y}}{\partial x_0} = -\frac{e}{\hbar c} A_x(x_0, y_0) + i \int d^2r g^* \frac{\partial g}{\partial x_0}
\end{equation}
The Berry connection finally reads:
\begin{equation}
\bm{A}_B = -\frac{e}{\hbar c} \bm{A}(x_0, y_0) + i \int d^2r g^* \frac{\partial g}{\partial x_0}
\end{equation}


\begin{thebibliography}{9}
\bibitem{berry1984}
M. V. Berry, ``Quantal phase factors accompanying adiabatic changes,'' \emph{Proceedings of the Royal Society of London. Series A, Mathematical and Physical Sciences}, vol. 392, no. 1802, pp. 45-57, Mar. 1984.
\bibitem{Konstantinou2016}
Georgios Konstantinou, Kyriakos Kyriakou, Konstantinos Moulopoulos,
\textit{Emergent Non-Hermitian Contributions to the Ehrenfest and Hellmann-Feynman Theorems},
\textit{IJEIR}, \textbf{5}(4), 2016, 248--252. ISSN: 2277-5668.
\bibitem{Konstantinou2018}
Georgios Konstantinou and Konstantinos Moulopoulos,
\textit{Topological anomalies in the off-diagonal Ehrenfest theorem and their role on optical transitions in solar cells},
\textit{Journal of Physics Communications}, \textbf{2}(8), 2018, 085011. 
\bibitem{RevModPhys.82.1959}
Di Xiao, Ming-Che Chang, and Qian Niu, 
``Berry phase effects on electronic properties,'' 
\textit{Reviews of Modern Physics}, vol.82, no.3, pp.1959--2007, Jul. 2010. 
\bibitem{kyriakou2020dynamicalextensionhellmannfeynmantheorem}
Kyriakos Kyriakou and Konstantinos Moulopoulos, 
\textit{Dynamical extension of Hellmann-Feynman theorem and application to nonadiabatic quantum processes in Topological and Correlated Matter}, 
arXiv:1506.08812 [cond-mat.mes-hall], 2020. 
\bibitem{Konstantinou2016}
G.~Konstantinou and K.~Moulopoulos,
``Generators of dynamical symmetries and the correct gauge transformation in the Landau level problem: use of pseudomomentum and pseudo-angular momentum''
\emph{European Journal of Physics}, vol.~37, no.~6, p.~065401, 2016.


\end{thebibliography}
\end{document}